\documentclass[fleqn,twoside]{article} % Specifies the document class
\usepackage{epsf,multicol,ifthen}%,mathptm}
\usepackage{ujp}
\usepackage[cp1251]{inputenc}
\usepackage[english,russian]{babel}
\usepackage{amstext}
\usepackage{amssymb}
\usepackage{cite}
\usepackage {graphicx}

\nazva{Dynamically broken symmetry in periodically gated quantum dots: Charge accumulation and dc-current }%
\udk{╣ ╙─╩/UDC }

\nazvacol{Dynamically broken symmetry}%

\avtor{T. Kwapi\'nski$^{1,2}$, S. Kohler$^{1,3}$ and  P. H{\"{a}nggi}$^{1}$ }%
\avtorcol{T. Kwapi\'nski, S. Kohler,  P. H{\"{a}nggi}}%

\inst{-}%
\adr{-}
\insti{{}Institute of Physics, University of Augsburg}%
\adri{(Univeristatsstr. 1, D-86135 Augsburg, Germany)}
\instii{{}Institute of Physics, Maria Curie-Sk\l odowska University}%
\adrii{(pl. M. Curie-Sk\l odowskiej 1, 20-031 Lublin, Poland) }
\instiii{Instituto de Ciencia de Materiales de Madrid, CSIC}%
\adriii{(C/Sor In\'es Juana de la Cruz 3, Cantoblanco, 28049 Madrid, Spain) }

\begin{document}           % End of preamble and beginning of text.
\setcounter{page}{1}%
\maketitl                 % Produces the title.
\begin{multicols}{2}
\anot{%
Time-dependent electron transport through a quantum dot and double
quantum dot systems in the presence of polychromatic external
periodic quantum dot energy-level modulations is studied within
the time evolution operator method for a tight-binding
Hamiltonian. Analytical relations for the dc-current flowing
through the system and the charge accumulated on a quantum dot are
obtained for the zero-temperature limit. It is shown that in the
presence of   periodic perturbations the sideband peaks of the
transmission are related to combination frequencies of the applied
modulations.  For a double quantum dot system under the influence
of polychromatic perturbations the quantum pump effect is studied
in the absence of source-drain and static bias voltages. In the
presence of spatial symmetry the charge is pumped through the
system due to
broken generalized parity symmetry.}%

%%%%%%%%%%%%%%%%%%%%%%%%%%%%%%%%%%%%%%%%%%%%%%%%%%%%%%%%%%%%%%%%%%%%
\section{\label{sec10}Introduction}
Recently, considerable progress has been achieved in fabricating
low dimensional systems and many experimental and theoretical
works have been put forward. Especially interesting are quantum
systems under the influence of external radio or microwave
electromagnetic radiation perturbations where many interesting
effects are observed like photon assisted tunneling (PAT)
\cite{Oost,Wiel}, turnstile effects and photon-electron quantum
pumps \cite{Sta,Kouw,Kouw2}, conductance oscillations \cite{Kwap},
and alike \cite{Koh2}.

The symmetry of quantum dot (QD) systems (with no source-drain voltage)
plays the crucial role as concerns electron pumping effects.
Generally, one can consider  symmetries like
time-reversal symmetry, time-reversal parity and generalized  parity \cite{Koh2}.
%The first one is conserved if an external perturbation, $a(t)$,
%satisfies the relation: $a(t) = a(-t)$, like e.g. cosine function,
%the second one if $a(t) = -a(-t)$, like e.g. sinus function, and
%the last one if $a(t) = -a(t+\pi/\omega)$, where $\omega$ stands
%for a frequency of external perturbation.

A  single electron pump based on asymmetrical couplings between QD
and the left and right electrodes was considered in Ref.
\cite{Kouw}. The couplings were switched on and off alternatively
from zero to maximal values (by means of additional electrodes)
and it led to electron pumping. A similar effect can be achieved
for dipole driving forces applied to a double QD system (in the
large gate voltage regime) or to quantum wires. In this case one
QD site is driven by the external dipole interaction which is out
of phase in comparison with the perturbation applied to the second
QD site (the QD sites are not driven in homogeneous way), e.g.
\cite{Cam,Koh2,platero,Koh3, arrachea,strass}. However, in the
presence of spatial symmetry (and in absence of a source drain and
static gate voltages) it is also possible to pump electrons but
the symmetry must be broken in a dynamical way. The easiest way to
break the time-reversal symmetry is to add a second harmonic to
the driving system; i.e.  a so called ``harmonic mixing'' drive
\cite{Koh2,Leh2,Mah}, or in general the second external
perturbation with an arbitrary frequency, \cite{Zhao}. In such a
case, depending on the parameters of these two time-dependent
perturbations perturbations  the generalized parity can be broken
and nonvanishing  current can flow through the system
\cite{Koh2,Leh2,RMPBM}.

There are few studies which address   the electron transport
through low-dimensional systems in the presence of several
polychromatic external perturbations with arbitrary frequencies.
Due to numerical problems, most of them concentrate on the case of
commensurate frequencies or only bi-harmonic perturbations. It was
shown that the external bi-harmonic time-dependent perturbations
can be used to control the noise level in quantum systems
\cite{Cam,Cam2} or, as well, for routing optically induced
currents \cite{Leh,Koh}. The shot noise for a single level quantum
dot under the influence of two ac external perturbations (coherent
or incoherent) was analyzed in Ref. \cite{Zhao}. The coherent
destruction of tunneling \cite{CDT} and the associated dynamical
localization in quantum dots under the influence of a
time-dependent perturbation with many harmonics were investigated
in Ref. \cite{Bas}. Moreover, dissipative quantum transport in one
or two dimensional periodic systems that are subjected to electric
harmonic mixing perturbations (bi-harmonic) were studied in Refs.
\cite{Goy,Sen,Bor}. The nonlinear signal consisted of e.g. two
rectangular-like driving forces which in turn allow to control
overdamped transport in Brownian motor devices,
\cite{RMPBM,Sav,Sav2,Sav3}.

In this paper we shall investigate the influence of polychromatic
time-dependent energetic perturbations with arbitrary
(commensurate and incommensurate) frequencies applied to the QD or
double QD system attached to leads for charge accumulated on the
QD and the time-averaged, dc-current flowing through the device.
For a double QD system we propose a quantum pump which is based on
a scheme which mimics closely  a dipole-like perturbation. Thus,
our work can be treated as a generalization of the studies of the
electron transport through a QD or double QD systems affected by
one external perturbation or bi-harmonic electric time-dependent
ac-perturbations with arbitrary frequencies. A tight-binding
Newns-Anderson Hamiltonian and evolution operator method are used
in our calculations.

The paper is organized as follows. In Sec. \ref{sec20} the model
Hamiltonian and theoretical description of a single level quantum
dot are presented. Also analytical relations for the time-averaged
current and time-averaged charge on the QD are obtained and
numerical results are depicted and interpreted. In Sec.
\ref{sec30} the current through a double QD system is obtained
and the pumping effect is discussed. The last section, Sec.
\ref{sec40}, is devoted to conclusions.

\section{\label{sec20} Single-Level Quantum Dot}

\subsection{\label{sec21}Theoretical description}

In this section, starting from the second quantization Hamiltonian
and using the evolution operator method we obtain the charge
accumulated on a QD and the  current flowing through the system
under the influence of many external time-dependent perturbations. Our system consists
of a single level quantum dot and two connecting electron electrodes, left
(L) and right (R). The total  Hamiltonian is then given by $H = H_0 +
V$, where
\begin{equation}
H_0 = \sum_{\vec k\alpha=L,R} \varepsilon_{\vec k\alpha} c^+_{\vec
k\alpha} c_{\vec k\alpha} + \varepsilon_d(t) c^+_{d} c_{d}\,,
\label{eq1}
\end{equation}
\begin{equation}
V = \sum_{\vec kL} V_{\vec kL} c^+_{\vec kL}c_d +
    \sum_{\vec kR} V_{\vec kR} c^+_{\vec kR} c_d+{\rm h.c.}
%    \sum^{N-1}_{i=1} V_{N} c^+_{i} c_{i+1} + {\rm h.c.}\,
%V = \sum_{\vec kL} (V_{\vec kL} c^+_{\vec kL}c_1  + {\rm h.c.}) +
%    \sum_{\vec kR} (V_{\vec kR} c^+_{\vec kR} c_N + {\rm h.c.}) +
%    \sum^{N-1}_{i=1} (V_{N} c^+_{i} c_{i+1} + {\rm h.c.})\,.
\label{eq2x}
\end{equation}
The operators $c_{\vec k\alpha}(c^+_{\vec k\alpha})$, $c_d(c^+_d)$
are the annihilation (creation) operators of the electron in the
lead $\alpha$ ($\alpha = L, R$) and at QD, respectively. The QD is
coupled symmetrically to the leads through the tunneling barriers
with the transfer-matrix elements $V_{\vec kL}$ and $V_{\vec kR}$
(hopping integrals). For the role of asymmetric lead-``molecule''
coupling see in Refs. \cite{Petrov1,Petrov2}. The
electron-electron Coulomb interaction is neglected in our
calculation, cf. \cite{Koh2,Zhao,Kwap2009}.

External perturbations are applied to the QD (the QD energy level
is driven in time by  time-dependent ac-voltages). We consider
here a harmonic modulation of the external energy level perturbations applied
to the QD, i.e.
\begin{equation}\label{eqed}
\varepsilon_d(t)=\varepsilon_d+\sum_{i=1}^n \Delta_i \cos(\omega_i
t+\phi_i)
\end{equation}
where $\omega_i$, $\Delta_i$ and $\phi_i$ are the frequency,
driving amplitude and phase of $i$-th perturbation.

The current flowing through the system and charge localized at QD
can be described in terms of the time evolution operator
$U(t,t_0)$ given by the equation of motion
 (in the interaction representation, $\hbar=1$), i.e.,
\begin{equation}
 i{\partial\over\partial t} U(t,t_0) = \tilde V(t)\,U(t,t_0)\,,
\label{eq3}
\end{equation}
where $\tilde V(t) = U_0(t,t_0) \, V(t) \, U^+_0(t,t_0) $ and
$U_0(t,t_0) = T\exp\left(i\int^t_{t_0} dt' H_0(t')\right)$. The
knowledge of the appropriate matrix elements of the evolution
operator $U(t,t_0)$ allows us to find the charge accumulated on the QD,
$n_d(t)$ (cf. \cite{Kwap,Gri,Tar}), which is given by:
\begin{eqnarray}
{n_d(t)}= \sum_{\beta}n_{\beta}(t_0)|U_{d,\beta}(t,t_0)|^2 \,, \,
\label{eq5}
\end{eqnarray}
where $n_{\beta}(t_0)$ represents the initial filling of the
corresponding single-particle states ($\beta=d,\vec kL, \vec kR$).
The current flowing e.g. from the left lead can be obtained from
the time derivative of the total number of electrons in the left
lead, cf. \cite{Jau}:
\begin{equation}
j_L(t) = -edn_L(t)/dt\,, \label{eq4}
\end{equation}
where $n_L(t)$ can be expressed as follows:
\begin{eqnarray}
{n_L(t)}=\sum_{\vec kL} n_{\vec kL}(t) = \sum_{\vec kL}
\sum_{\beta}n_{\beta}(t_0)|U_{\vec kL,\beta}(t,t_0)|^2 \,. \,
\label{eq5b}
\end{eqnarray}
Using Eq.~\ref{eq3} the following differential equations for
$U_{d, \beta}(t,t_0)$ and $U_{\vec kL, \beta}(t,t_0)$ matrix
elements which are needed to obtain the current and QD charge, can
be written in the form:
\begin{eqnarray}
i{\partial\over\partial t} U_{d, \beta}(t,t_0) = \sum_{\vec k
\alpha=L,R} \tilde V_{d, \alpha}(t) U_{\alpha, d} (t,t_0)
\label{eq6}
\end{eqnarray}
\begin{eqnarray}
i {\partial\over\partial t} U_{\vec kL, \beta}(t,t_0) =\tilde
V_{\vec kL, d}(t) U_{d, \beta} (t,t_0)
 \label{eq6b}
\end{eqnarray}
where nonzero elements of the function $\tilde V$ are
\begin{eqnarray}
\tilde V_{d,\vec k \alpha} =\tilde V^*_{\vec k\alpha,d}=V^*_{\vec
k\alpha} \exp{\left(i \int_{t_0}^t
(\varepsilon_d(t')-\varepsilon_{\vec k\alpha} )dt'\right)}
\label{eq6c}
\end{eqnarray}

Next, using the {\it wide band limit} approximation
$\Gamma^{\alpha}(\varepsilon)=2\pi \sum_{\vec k\alpha}V_{\vec
k\alpha} V^*_{\vec k\alpha}\delta(\varepsilon-\varepsilon_{\vec
k\alpha})=\Gamma^{\alpha}$ and assuming $\Gamma^L=\Gamma^R=\Gamma$
one can find the following relation for $U_{d,d}(t)$ matrix
element ($t_0=0$): $U_{d,d}(t) = \exp{(-\Gamma t)}$, and similar
for $U_{d, \vec k\alpha}(t)$ and $U_{\vec kL, \beta}(t)$:
\begin{eqnarray}\label{2x2}
U_{d, \vec k\alpha}(t) &=&-i \exp{(-\Gamma t)} \int_{0}^t dt'
\tilde V_{d, \vec k\alpha}(t') \exp{(\Gamma t')}
\end{eqnarray}
\begin{eqnarray}
U_{\vec kL, \beta}(t) &=&-i \int_{0}^t dt' \tilde V_{\vec
kL,d}(t')  U_{d, \beta}(t')
\end{eqnarray}
Note,  that for $t \rightarrow \infty$ the element $U_{d,d}(t)$
tends to zero and thus the charge accumulated on the  QD does not depend
on the initial QD occupation $n_{\beta}(t_0)$ for large $t$. Also the current
flowing through the system is independent on $n_d(t_0)$ because
the element ${d \over dt} \sum_{\vec kL} n_d(t_0) |U_{\vec
kL,d}(t)|^2$  tends to zero for $t \rightarrow \infty$. Finally,
the QD charge can be written in the form:
\begin{eqnarray}\label{x1}
{n_d(t)} &=& \sum_{\vec k \alpha =L,R} n_{\vec k
\alpha}(t_0)|U_{d,\vec k \alpha}(t)|^2
\end{eqnarray}
and the current through the system reads ($e=1$):
\begin{eqnarray}\label{x2}
j_L(t)&=&\Gamma n_d(t)+Im \left( \sum_{\vec k L} n_{\vec kL}(t_0)
\tilde V_{\vec kL, d}(t) U_{d, \vec kL} (t) \right) \nonumber\\
\end{eqnarray}
Eq.~\ref{x1} and Eq.~\ref{x2} are very general relations which
should be analyzed using Eq.~\ref{eq6c} and Eq.~\ref{2x2}. The
relation for the current, Eq.~\ref{x2}, has the structure which
can be written by means of the transmission
$T_{LR}$ and $T_{RL}$, i.e. $j_L(t)= \sum_{\vec k L}n_{\vec k
L}(0) T_{LR}(t)- \sum_{\vec k R}n_{\vec k R}(0) T_{RL}(t)$, but note that in
general $T_{LR}(t)\neq T_{RL}(t)$. In order to obtain the current
one should know the exact form of $\tilde V_{d,\vec k \alpha}$
function. Using the time-dependence relation for the QD-energy
level, Eq.~\ref{eqed}, and assuming $\phi_i=0$, the elements
$\tilde V_{d,\vec k \alpha}$, Eq.~\ref{eq6c}, can be expressed as
follows:

\begin{eqnarray}\label{x16}
\tilde V_{d,\vec k \alpha} =V^*_{\vec k\alpha} e^{i
(\varepsilon_d-\varepsilon_{\vec k\alpha})t} \prod_{i=1}^n
\sum_{m_i} J_{m_i}\left({\Delta_i \over \omega_i} \right)
e^{im_i\omega_it}
\end{eqnarray}
and the solution of Eq.~\ref{2x2} for the evolution operator
elements can be written in the form:
\begin{eqnarray}
U_{d, \vec k\alpha}(t) &=&-V_{\vec k\alpha}^*
e^{i(\varepsilon_d-\varepsilon_{\vec k \alpha}) t}\times
\nonumber\\&& \sum_{m_1}...\sum_{m_n}
 {J_{m_1}\left({\Delta_1 \over \omega_1}
\right)...J_{m_n}\left({\Delta_n \over \omega_n} \right) e^{i
\Omega t} \over{\varepsilon_d-\varepsilon_{\vec k \alpha} +\Omega
-i \Gamma }}
\end{eqnarray}
where $m_i \in (-\infty, \infty)$ and is an integer number,
$\Omega=m_1 \omega_1+...+m_n \omega_n$ and $J_i$ is the Bessel
function. Next, we obtain the dc current  through the system,
$j_0=\langle j(t)\rangle=\lim_{T\rightarrow \infty}{1 \over T}
\int_{-T/2}^{T/2} j(t')dt'$, which can be symmetrized, $j_0
=\langle j_L(t)\rangle=\left( \langle j_L(t)\rangle-\langle j_R(t)
\rangle \right)/2$, and finally we find:
\begin{eqnarray}\label{t00}
j_0 &=& {\Gamma \over 2 \pi} \int_{-\infty}^\infty \left(
f_R(\varepsilon)- f_L(\varepsilon) \right) T(\varepsilon)
\end{eqnarray}
where $f_{L/R}(\varepsilon)$ is the Fermi function of $L/R$
electrode and the transmission reads
\begin{eqnarray}\label{t01}
T(\varepsilon)&=&\Gamma^2 \sum_{m_1}...\sum_{m_n}
\sum_{m_1^{'}}...
\sum_{m_n^{'}} \delta_{\Omega-\Omega^{'}} \times \\
&&{{{J_{m_1}\left({\Delta_1 \over \omega_1}
\right)...J_{m_n}\left({\Delta_n \over \omega_n}
\right)J_{m_1^{'}}\left({\Delta_1 \over \omega_1} \right) ...
J_{m_n^{'}}\left({\Delta_n \over \omega_n} \right)}} \over
(\varepsilon_d-\varepsilon+\Omega)^2+\Gamma^2}  \nonumber
\end{eqnarray}
where $\Omega^{'}=m_1^{'} \omega_1+...+m_n^{'} \omega_n$. Note,
that for incommensurate frequencies the Kronecker Delta function
$\delta_{\Omega-\Omega^{'}}$ is nonzero only for $m_i^{'}=m_i$ and
thus the transmission can be written in the following short form:
\begin{eqnarray}\label{t02}
T(\varepsilon)= \Gamma^2 \sum_{m_1}...\sum_{m_n}
{J_{m_1}^2\left({\Delta_1 \over \omega_1}
\right)...J_{m_n}^2\left({\Delta_n \over \omega_n} \right) \over
(\varepsilon_d-\varepsilon+\Omega )^2+\Gamma^2}
\end{eqnarray}

This equation is valid also for the case when there is large
difference between frequencies $\omega_i$. Note, that the
transmission, (Eqs.~\ref{t01} and \ref{t02}), corroborates with
the results obtained by means of Green's function method for a
quantum wire driven by homogeneous external perturbations,
\cite{Kwap2009}.

The relation for the current, Eq.~\ref{t00}, has the structure of
the Landauer formula. We note that the structure
in Eq. (17) involves, in clear contrast to a electric field dipole perturbation \cite{Koh2},
 no inelastic photon-assisted tunneling events.
This is so because with the time-dependent energy level perturbation used  here 
the long  time-average of $T_{LR}(t)$ and
$T_{RL}(t)$ equals $T_{LR}(\epsilon)= T_{RL}(\epsilon)$. For  zero
temperature and for incommensurate frequencies of external
perturbations the dc-current can be obtained analytically from the
relation:

\begin{eqnarray}\label{t03}
j_0 &=& \Gamma \sum_{m_1}...\sum_{m_n} J_{m_1}^2\left({\Delta_1
\over \omega_1} \right)...J_{m_n}^2\left({\Delta_n \over \omega_n}
\right) \times
\\ && \left( \arctan{\varepsilon_d-\mu_L-\Omega \over \Gamma}
-\arctan{\varepsilon_d-\mu_R-\Omega \over \Gamma} \right)
\nonumber
\end{eqnarray}
Note that for $\mu_L=\mu_R$ the dc current is zero.

The corresponding analytical relation for
the accumulated quantum dot dc-charge; i.e.
$n_0=\langle n_{d}(t)\rangle=\lim_{T\rightarrow \infty}{1
\over T} \int_{-T/2}^{T/2} n_{d}(t')dt'$ reads:
\begin{eqnarray}\label{t04}
n_0 &=& {1 \over 2\pi} \sum_{m_1}...\sum_{m_n}
J_{m_1}^2\left({\Delta_1 \over \omega_1}
\right)...J_{m_n}^2\left({\Delta_n \over \omega_n} \right) \times
\\ && \left({\pi}- \arctan{\varepsilon_d-\mu_L-\Omega \over
\Gamma}- \arctan{\varepsilon_d-\mu_R-\Omega \over \Gamma} \right)
\nonumber
\end{eqnarray}
It is worth noting that for $\varepsilon_d << \mu$ ($\varepsilon_d
>> \mu$) the charge accumulated on the QD is maximal (minimal).
The relations for the dc current,
Eq.~\ref{t03}, and for the QD charge, Eq.~\ref{t04}, constitute the main
analytical relations of this section.

\subsection{\label{sec23}QD accumulated charge and dc current}

In this section we analyze the QD charge  and the dc current
flowing through a quantum dot driven by polychromatic
perturbations. All energies are expressed in the units of
$\Gamma^0$ and in order to obtain rather narrow sidebands peaks we
assume $\Gamma=0.2\Gamma^0$ (taking the unit of energy
$\Gamma^0=0.05eV$ it corresponds to $\Gamma=0.01eV$). For larger
$\Gamma$ all sideband peaks are wider and it is  difficult to
observe many-perturbation effects. The current and the conductance are
given in units of $2e\Gamma^0/\hbar$ and $2e^2/\hbar$,
respectively. Moreover, we show numerical calculations for two
external perturbations case but the generalization for more perturbations is
obvious.

In Fig.~1 the QD charge (upper panel) and the dc current flowing
through the system (lower panel) are shown for two external
perturbations applied to the system ($\omega_1=3$, $\omega_2=8$,
$\Delta_1=4$, $\Delta_2=8$) - thick lines. The frequencies are
commensurate but there is large difference between them and the
transmission obtained form Eq.~\ref{t01} and Eq.~\ref{t02} are
almost the same. Physically it means that eight-photon
adsorption/emission process based on the third sideband peak
should occur to play the role in the transmission,
($8\omega_1=3\omega_2$), but this is unlikely process. The broken
lines correspond to the single external perturbation case, i.e.
$\Delta_2=8$ ($\Delta_1=0$) - thin broken lines and $\Delta_1=4$
($\Delta_2=0$) - thick broken lines, respectively. The chemical
potentials are rather small, $\mu_L=-\mu_R=0.1$, and the current
peak for $\varepsilon_d=0$ (lower panel) is observed for mono and
polychromatic cases (it appears also for the time-independent
case).
%%%%%%%%%%%%%%%%%%%%%%%%%%%%%%%%%%%%%%%%%%%%%%%
\begin{center} \noindent \epsfxsize=0.9\columnwidth\epsffile{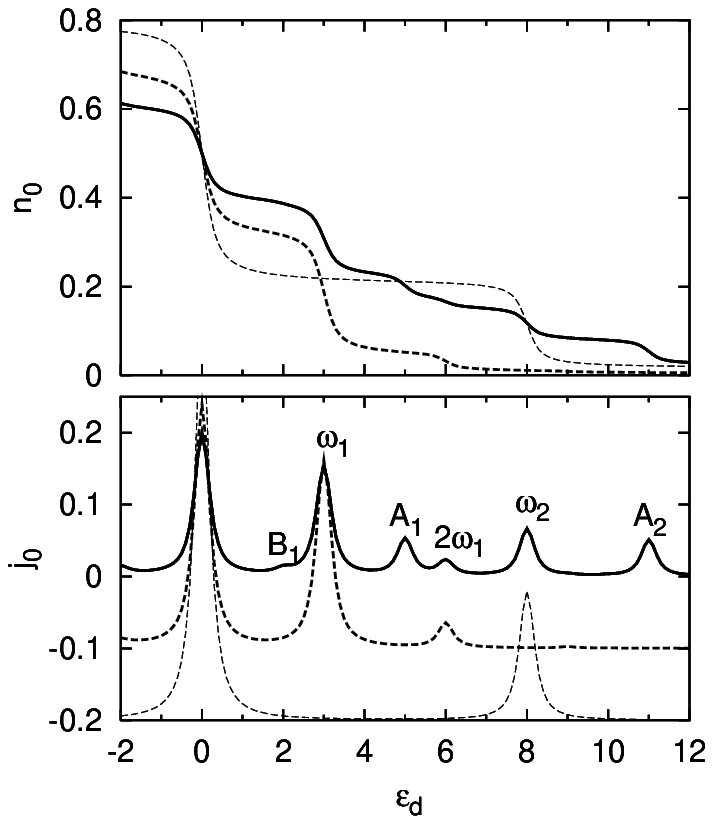}
\end{center}
\vskip-3mm\noindent{\footnotesize Fig. 1. QD charge (upper panel)
and the dc current (lower panel) as a function of $\varepsilon_d$
for $\Delta_1=4$ and $\Delta_2=8$ (thick lines), $\Delta_1=4$,
$\Delta_2=0$ (thick broken lines) and $\Delta_1=0$ and
$\Delta_2=8$ (thin broken lines), respectively. The other
parameters are $\omega_1=3$, $\omega_2=8$, $\mu_L=-\mu_R=0.1$,
$\Gamma=0.2$. The thick (thin) broken line on the lower panel is
shifted by -0.1 (-0.2) for better visualization. }%
\vskip15pt
%%%%%%%%%%%%%%%%%%%%%%%%%%%%%%%%%%%%%%%%%%%%%%%%
\noindent For only one external perturbation applied to the QD,
$\omega=3$ (or $\omega=8$), the sidebands peaks are visible for
$\varepsilon_d=\pm k \omega$, where $k$ is an integer number.
However, for the case when two external perturbations are applied
simultaneously to the QD, additional sideband peaks appear. Of
course the single peaks from the first and the second fields  are
still visible i.e. for $\varepsilon_d=\omega_1, 2\omega_1$ or
$\omega_2$. In Fig.~1, additional dc current peaks are indicated
by points $A_1$,  $A_2$ (first-order sidebands) and $B_1$
(second-order sideband). $A_1$ ($A_2, B_1$) sideband peak appears
for $\varepsilon_d=\omega_2-\omega_1$
($\varepsilon_d=\omega_2+\omega_1$,
$\varepsilon_d=\omega_2-2\omega_1$) and is related to the peak for
$\varepsilon_d=\omega_2=8$ (but not to the main peak for
$\varepsilon_d=0$). It is worth noting that although the
frequencies $\omega_1$ and $\omega_2$  are not equal to 5 or 11 we
observe the sideband peaks for these values of $\varepsilon_d$. In
general we observe sideband peaks for $\varepsilon_d=\pm k_1
\omega_1 \pm k_2 \omega_2$, where $k_1$ and $k_2$ are integer
numbers. The structure of the dc current curves are reflected also
in the charge accumulated on the QD (upper panel). The charge
decreases with $\varepsilon_d$ but there are many steps which are
related to the current sideband peaks. Also additional sidebands
i.e. points $A_1$, $A_2$ and $B_1$ from the lower panel, are
visible on the charge curve (Fig.~1, upper panel, thick line).

%\subsection{\label{sec34}QD charge and current}
%%%%%%%%%%%%%%%%%%%%%%%%%%%%%%%%%%%%%%%%%%%%%%%
\begin{center} \noindent \epsfxsize=0.9\columnwidth\epsffile{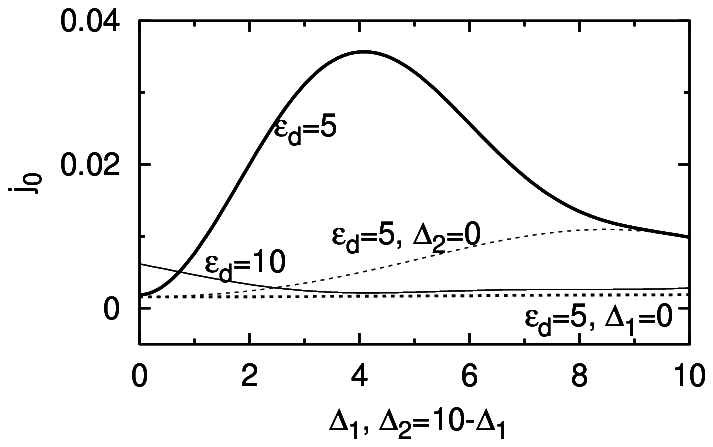}
\end{center}
\vskip-3mm\noindent{\footnotesize Fig. 2. The dc current as a
function of the amplitude $\Delta_1$ ($\Delta_2=10-\Delta_1$) for
$\varepsilon_d=5$ (thick line) and $\varepsilon_d=10$ (thin line),
$\omega_1=3$, $\omega_2=8$. The thin (thick) broken line
corresponds to one external perturbation case $\Delta_1$,
$\omega=3$ with $\Delta_2=0$ ($\Delta_2$, $\omega=8$
with $\Delta_1=0$) and $\varepsilon_d=5$; $\mu_L=-\mu_R=0.1$, $\Gamma=0.2$.}%
\vskip15pt
%%%%%%%%%%%%%%%%%%%%%%%%%%%%%%%%%%%%%%%%%%%%%%%%
Next, in  Fig. 2 we study the dc current as a function of the
driving strengths (amplitudes) of two external perturbations
applied to the system. For the second signal the driving strength
decreases with the amplitude of the first perturbation, i.e.
$\Delta_2=10-\Delta_1$. The thick (thin) solid line corresponds to
$\varepsilon_d=5$ ($\varepsilon_d=10$). As one can see for
$\varepsilon_d=10$ the dc current is very small and almost
independent on the amplitudes $\Delta_1$ and $\Delta_2$. This
conclusion is valid for every $\varepsilon_d$ which is not any
combination of $\omega_1$ and $\omega_2$. For $\varepsilon_d=5$
($\varepsilon_d=\omega_2-\omega_1$) the current is minimal for
$\Delta_1=0$ and $\Delta_1=10$ but for the mixed regime of
$\Delta_1$ and $\Delta_2$ it is characterized by a local maximum,
cf. thick solid line. This is a very interesting effect because one
can  control the current flowing through the system by applying
additional time-dependent perturbation. Note, that the maximal
value of the current is a few times larger than for the case of
one external perturbation, i.e. for $\Delta_1=0$ or $\Delta_1=10$.
For only one external perturbation applied to the system the dc
current is shown by the broken lines - the thin line for
$\Delta_2=0$ as a function of $\Delta_1$ ($\omega=3$) and the
thick one for $\Delta_1=0$ as a function of $\Delta_2$
($\omega=8$), $\varepsilon_d=5$. One can conclude that one
external perturbation slightly changes the current in this case.
It confirms that the current maximum, which appears for
$\varepsilon_d=5$ (thick solid curve), is due to a combination of
two external perturbations applied to the system. The similar
conclusions are valid for other $\varepsilon_d=\pm k_1 \omega_1\pm
k_2 \omega_2$ ($k_{1,2}\neq 0$).

%%%%%%%%%%%%%%%%%%%%%%%%%%%%%%%%%%%%%%%%%%%%%%%
\begin{center} \noindent \epsfxsize=0.9\columnwidth\epsffile{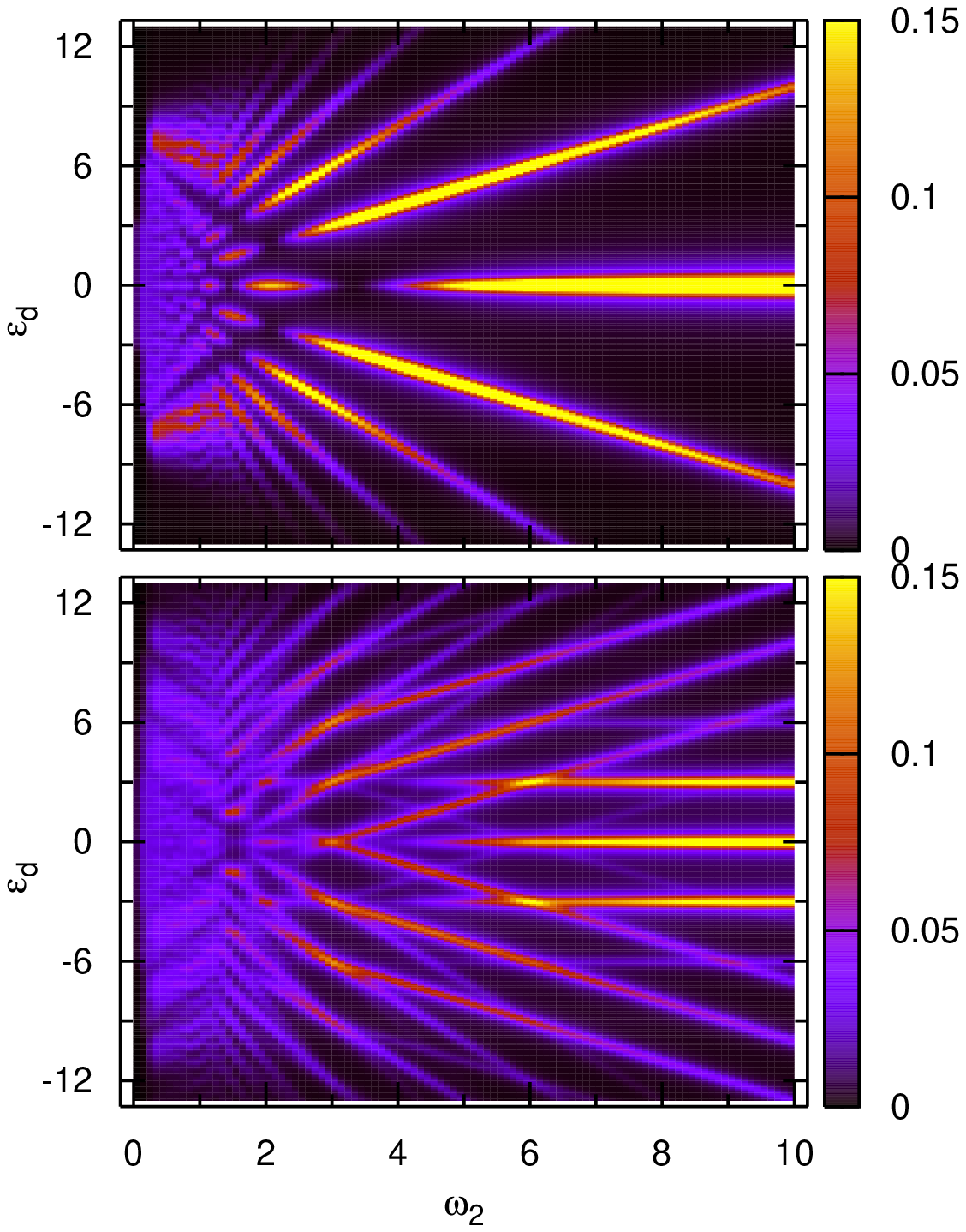}
\end{center}
\vskip-3mm\noindent{\footnotesize Fig. 3. (Color online) The dc current as a
function of $\varepsilon_d$ and $\omega_2$ for $\Delta_1=0$,
$\Delta_2=8$ (one external perturbation - upper panel) and
$\Delta_1=3$, $\Delta_2=8$ (two external perturbations - lower
panel). The other parameters are:  $\omega_1=3$, $\mu_L=-\mu_R=0.1$, $\Gamma=0.2$.}%
\vskip15pt
%%%%%%%%%%%%%%%%%%%%%%%%%%%%%%%%%%%%%%%%%%%%%%%%
To analyze the structure of sidebands peaks for polychromatic
perturbations in Fig.~3 we show the dc current as a function of
the QD energy level, $\varepsilon_d$, and the frequency of
external perturbation $\omega_2$ for $\Delta_1=0$ (only one
perturbation applied to the system with $\Delta_2=8$) - upper
panel, and for two external perturbations with $\Delta_1=3$,
$\omega_1=3$, $\Delta_2=8$ - lower panel. For one, as well as for
two time-dependent perturbations applied to the system we can
distinguish between the adiabatic regime for $\omega_2\leq 1$ and
the non-adiabatic one for $\omega_2>1$. For $\omega_2>1$ one can
observe the main current peak for $\varepsilon_d=0$ and sidebands
which are visible for $\varepsilon_d=\pm k \omega_2$. The
situation is more complex for two external perturbations applied
simultaneously to the QD ($\Delta_1=3$, $\omega_1=3$,
$\Delta_2=8$). One observes  that each light line from the upper
panel possesses satellite lines which are localized in the
distance $\pm k \omega_1$ from the original lines. Thus we observe
very reach structure of the dc current in the presence of
polychromatic perturbations.

\subsection{\label{sec24}Results: Transmission asymmetry and phase dependence }

For one external perturbation  applied to a QD  as well as for two
external perturbations with the same frequencies one observes
fully symmetric current curves around the value $\varepsilon_d=0$.
However, for two external time-dependent perturbations the dc
current (or the transmission)  can be asymmetrical which appears
for the case of commensurate frequencies. In Fig. 4 we show the dc
current obtained for the frequency ratio $\omega_2/\omega_1=2, 1$
and $0.5$ using the transmission relations, Eq.~\ref{t01}, (solid
lines) and for the case where there is an infinitesimal shift of
the frequency ratio from an integer number,  Eq.~\ref{t02} (broken
lines), e.g. for $\omega_2/\omega_1=2$ we set $\omega_1=3$ and
$\omega_2=5.999$. Note, that all broken curves in Fig. 4 are
symmetrical. The solid lines, however, are asymmetrical around
$\varepsilon_d=0$ except for the case $\omega_1$ = $\omega_2$.
Notably, for $\omega_1=\omega_2$ the system is equivalent to  a
single, effective external perturbation case with the effective
driving amplitude, $\Delta=\Delta_1+\Delta_2$, and thus the
current curves render themselves symmetrical again. The current
obtained for incommensurate frequencies differs from the ones of
commensurate frequencies. It is worth noting that the structure of
the dc current curves for fixed $\omega_i$ depends on the
amplitudes of external perturbations but the positions of peaks
remain unchanged.
%%%%%%%%%%%%%%%%%%%%%%%%%%%%%%%%%%%%%%%%%%%%%%%
\begin{center} \noindent \epsfxsize=0.88\columnwidth\epsffile{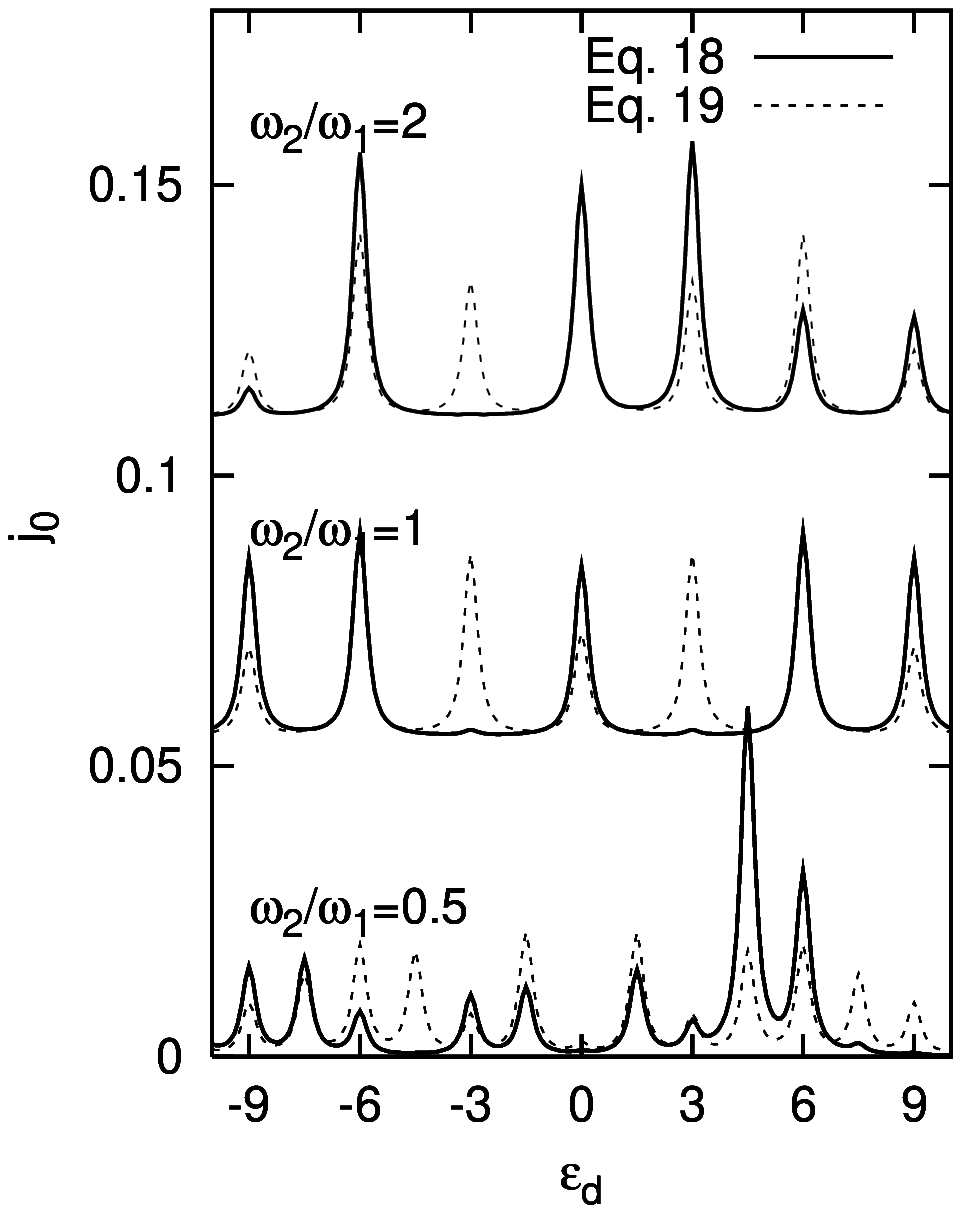}
\end{center}
\vskip-3mm\noindent{\footnotesize Fig. 4. The dc current as a
function of $\varepsilon_d$ obtained for commensurate frequency
ratios $\omega_2/\omega_1=2, 1$ and $0.5$ according to
Eq.~\ref{t00} with the transmission given by Eq.~\ref{t01} (solid
lines) and for a slightly off-commensurate frequency; i.e.,
$\omega_1 = 3, \omega_2 = 5.999$, see Eq.~\ref{t02}  (broken
lines).
All parameters are the same as in Fig.~3.}%
\vskip15pt
%%%%%%%%%%%%%%%%%%%%%%%%%%%%%%%%%%%%%%%%%%%%%%%%

Next, it is necessary to explain the transmission and current
asymmetries for the case of commensurate frequencies. In this case
to obtain the transmission one should use Eq.~\ref{t01}  which is
expressed by means of four Bessel functions, an energy dependent
factor and the Kronecker Delta function. For commensurate frequencies this
Delta function produces off-diagonal elements of the Bessel
functions which are multiplied by the energy factor. The energy
factor depends on the order of the Bessel functions and introduces
asymmetry for positive and negative values of $\varepsilon_d$ (it
means that the energy factor possesses different values for
positive and negative integer order of the Bessel functions). The
off-diagonal elements of the Bessel functions are also responsible
for different values of the dc current obtained for
$\omega_2/\omega_1=1$, according to Eq.~\ref{t01} and
Eq.~\ref{t02}. It is worth noting that in our calculation we
assume that the QD energy level is driven  according to the
formula with the cosine function (i.e. the cosine function is an odd function) and
the external time-dependent perturbation satisfies the
time-reversal symmetry, $a(t)=a(-t)$.
%%%%%%%%%%%%%%%%%%%%%%%%%%%%%%%%%%%%%%%%%%%%%%%
\begin{center} \noindent \epsfxsize=0.9\columnwidth\epsffile{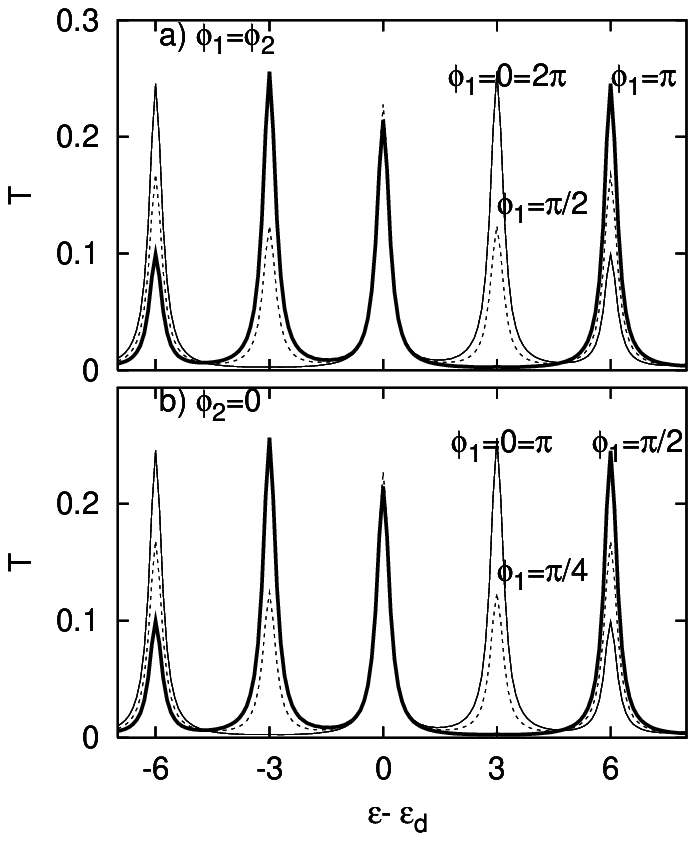}
\end{center}
\vskip-3mm\noindent{\footnotesize Fig. 5. The transmission versus
the energy for two external perturbations applied to the QD
($\omega_1=3$, $\omega_2=6$, $\Delta_1=3$, $\Delta_2=8$) with
phases: $\phi_1=\phi_2=0, \pi/2, \pi$ (upper panel) and $\phi_1=0,
\pi/4, \pi/2$ and $\phi_2=0$ (lower panel). }%
\vskip15pt
%%%%%%%%%%%%%%%%%%%%%%%%%%%%%%%%%%%%%%%%%%%%%%%%
\noindent In Fig.~5 we show the transmission for nonzero phase
factors of the external perturbations i.e. $\phi_1$ and $\phi_2$,
cf. Eq.~\ref{eqed}. The upper panel shows the results for the case
when both phases are changed in the same way i.e. $\phi_1=\phi_2$.
For $\phi_1=0$ (thin solid line) the asymmetry in the transmission
is observed, cf. also Fig.~4. If both phases are equal to $\pi/2$
(broken line) the transmission is symmetrical (as a function of
the energy) - in that case the external signals are sinuses. For
$\phi_1=\pi$ the transmission is asymmetrical again. The period of
the external perturbation for the case of the same phases,
$\phi_1=\phi_2$, is equal to $2\pi$. The situation is somewhat
different when the phase of one external perturbation is constant
($\phi_2=0$ - lower panel). Here, only for $\phi_1=\pi/4$ the
transmission  is symmetrical but for $\phi_1=0$ and $\pi/2$ this
function is asymmetrical. Note, that the above conclusions are
valid for the case of commensurate frequencies. For incommensurate
frequencies the transmission curves are symmetrical, cf. Fig. 4.
Moreover, as one can see, the transmission curves in the upper and
lower panels are the same, only the phases of external
perturbations are different. It means that the same effect can be
obtained by changing only one phase parameter  instead of driving
both phases simultaneously.

\section{\label{sec30}Double quantum dot system and pumping effect}

The asymmetry effect in the transmission obtained in the previous
section can be used to construct a single electron pump
\cite{strass} based on a two-level fully symmetrical system with
no source-drain and static bias voltages applied to the system. In
this section we analyze the electron transport through a double
quantum dot in the presence of external perturbations. The
Hamiltonian of our system can be written as follows, in close
analogy to Eq.~\ref{eq1} and Eq.~\ref{eq2x}; i.e.,

\begin{equation}
H_0 = \sum_{\vec k\alpha=L,R} \varepsilon_{\vec k\alpha} c^+_{\vec
k\alpha} c_{\vec k\alpha} + \varepsilon_1(t) c^+_{1} c_{1}
+\varepsilon_2(t) c^+_{2} c_{2}\,, \label{eq1x}
\end{equation}
\begin{equation}
V = \sum_{\vec kL} V_{\vec kL} c^+_{\vec kL}c_1 +
    \sum_{\vec kR} V_{\vec kR} c^+_{\vec kR} c_2+V_{12} c^+_{1} c_{2} +  {\rm h.c.}
\label{eq2}
\end{equation}
where $V_{12}$ is the tunnel coupling (hopping term) between two QD
sites. As before, we assume that there are only two external
perturbations applied to the system and one can write the following time
dependence of the quantum dot energy levels:
\begin{eqnarray}\label{eqed2a}
\varepsilon_1(t)=\varepsilon_1+\Delta_1 \cos(\omega_1 t)+\Delta_2
\cos(\omega_2 t)
\end{eqnarray}
\begin{eqnarray} \label{eqed2b}
 \varepsilon_2(t)=\varepsilon_2+\Delta_1
\cos(\omega_1 t+\phi)+\Delta_2 \cos(\omega_2 t+\phi)
\end{eqnarray}
where $\phi$ is the phase difference between the external perturbations
applied to the first and the second QD sites which play a similar
role as dipole forces in one external harmonic field case, cf.
\cite{Cam,Koh2,Koh3}.
We stress, however, that here the driving is chosen uniform in the sense that the amplitudes strengths
 for the driving of the two dots are identical. Note, that the role of phase difference
between the first and the second external perturbation was
considered in \cite{Leh2} for bi-harmonic perturbations and here
we concentrate on  the phase difference between both QD sites. The
main reason we omit the phase difference between both external
perturbations is that in the presence of spatial symmetry
($\varepsilon_1=\varepsilon_2$, $\mu_L=\mu_R$; i.e., in absence
of a static gate voltage and a  source-drain voltage), it is
impossible to pump electrons through the system.  By introducing
the phase difference according to Eq.~\ref{eqed2a} and
Eq.~\ref{eqed2b}, we change the symmetry  at the first and the
second QD sites and thus electrons can be pumped in the presence
of external perturbations and spatial symmetry. The general
formula for the time-dependent current flowing from the left
electrode can be written as follows:

\begin{eqnarray}\label{dqd01}
j_L(t)&=& \sum_{\vec k L} n_{\vec kL}(t_0) \left(  \Gamma |U_{1,
\vec kL}(t)|^2+ Im   \tilde
V_{\vec kL, 1}(t) U_{1, \vec kL} (t)  \right) \nonumber\\
&+& \sum_{\vec k R} n_{\vec kRL}(t_0) \Gamma |U_{1, \vec kR}
(t)|^2
\end{eqnarray}
 where $V_{\vec kL, 1}(t) $ is defined by Eq.~\ref{x16} and
the evolution operator matrix elements satisfy the following set
of differential equations (and similar for $U_{1(2),\vec kR}$):
\begin{eqnarray}\label{dqd02a}
{\partial\over\partial t} U_{1,\vec kL}(t) &=& -i V_{12} e^{i
(\varepsilon_1-\varepsilon_2)t} e^{i (f_1-f_2)} U_{2,\vec kL}(t)
\nonumber\\ && -i V_{\vec kL} e^{i
(\varepsilon_1-\varepsilon_{\vec kL})t} e^{i f_1}-{\Gamma \over 2}
U_{1,\vec kL}(t)
\end{eqnarray}
\begin{eqnarray}\label{dqd02}
{\partial\over\partial t} U_{2,\vec kL}(t) &=&-i V_{12} e^{i
(\varepsilon_2-\varepsilon_1)t} e^{i (f_2-f_1)} U_{1,\vec kL}(t)
\nonumber\\ &&-{\Gamma \over 2} U_{2,\vec kL}(t)
\end{eqnarray}
where $f_1={\Delta_1 \over \omega_1} \sin(\omega_1 t)+{\Delta_2
\over \omega_2} \sin(\omega_2 t)$, $f_2={\Delta_1 \over \omega_1}
\sin(\omega_1 t+\phi)+{\Delta_2 \over \omega_2} \sin(\omega_2
t+\phi)$. Note, that for  $\phi= 2\pi k$, and the same on-site
energies $\varepsilon_1=\varepsilon_2=\varepsilon_d$, one can
obtain the dc current and the transmission analytically. In this
case the current satisfies the Landauer formula and the
time-averaged left-right and right-left transmissions are equal.
This result is due to the above mentioned chosen uniform driving
strengths.
 For incommensurate frequencies of the external
perturbations the formula for the transmission simplifies and
can be written as follows:
\begin{eqnarray}\label{dqd03}
T(\varepsilon) &=& \Gamma^2 \sum_{m_1} \sum_{m_2} V_{12}^2 \times
\\&& {J_{m_1}^2\left({\Delta_1 \over \omega_1}
\right)J_{m_2}^2\left({\Delta_2 \over \omega_2} \right) \over
\left( (\varepsilon_d-\varepsilon+\Omega )^2-{\Gamma \over
2}^2-V_{12}^2 \right)^2 +\Gamma^2
(\varepsilon_d-\varepsilon+\Omega )^2  }\nonumber
\end{eqnarray}
where $\Omega=\omega_1 m_1+\omega_2 m_2$.  For the case of
commensurate frequencies the formula for the transmission assumes
no simple transparent from, but is similar to Eq.~\ref{t01}. Note,
that also for $\phi=0$ and $\varepsilon_1 \neq \varepsilon_2$ it
is possible to analyze the set of differential equations
analytically, Eq.~\ref{dqd02a} and Eq.~\ref{dqd02}, however the
solution is rather complex. Generally, i.e. for $\phi \neq 0$ and
$\varepsilon_1 \neq \varepsilon_2$ we solve the set of
differential equations for $U_{1(2),\vec kL(R)}$ numerically, then
put the solution into the relation for the current,
Eq.~\ref{dqd01}, and time-average it over the common period of
external perturbations. Thus, this procedure can be applied only
for the case of commensurate frequencies. In our calculations we
concentrate mainly on the phase difference effect as it can lead
to the electron pumping in the system with no source-drain and
static bias voltages. We take into consideration phase difference,
$\phi$, between QD sites and for $\phi=\pi/2$ it leads to a
dipole-like parameterization (the oscillations are out of phase).
The similar two-level system under the influence of one external
perturbation for the case of high bias voltage, $\varepsilon_1 <<
\varepsilon_2$, was investigated in Ref. \cite{Kai} and in the
presence of phase difference between both external perturbations
and dipole driving forces in Ref. \cite{Leh2}.

If only one external perturbation is applied to the double dot
system, the phenomenon electron pumping does not occur for the
case of spatial symmetry $\varepsilon_1 = \varepsilon_2$,
\cite{Koh2}. For nonzero gate voltage and $\phi=\pi/2$ (i.e. in
the presence of the phase and spatial asymmetry) the current can
flow through the system, Ref. \cite{Kai}, but if there is no phase
difference between both QD sites, $\phi=0$, the current vanishes
again. In order to analyze the role of polychromatic perturbations
on the symmetrical system i.e. $\varepsilon_1 = \varepsilon_2=0$
(no applied static gate voltages) we depict in Fig.~6 the dc
current  as a function of the driving amplitude, $\Delta_2$, for
two external signals and for the phase difference between QD sites
$\phi=0=2\pi$ (thick broken line), $\phi=\pi$ (thin broken line)
and $\phi=\pi/2$ (solid line). For $\Delta_2=0$ the current is
zero, independent on the phase $\phi$; in this case   only one
external perturbation is acting and no symmetry breaking taking
place. For two external perturbations, and for no phase difference
between  the first and the second QD sites, the current also does
not flow, being equal to zero for all $\Delta_2$ (thick broken
line). However, for $\phi\in (0,2\pi) $ the current flows although
there is no voltages applied to the system. In that case, we
realize a sort of a quantum pump which works only in the presence
of two time-dependent perturbations, implying that generalized
parity is broken in a dynamical way \cite{Koh2,Leh2}. Note, that
depending on the phase difference between both QD sites the dc
current can be positive or negative, cf. the broken and solid
lines.
%%%%%%%%%%%%%%%%%%%%%%%%%%%%%%%%%%%%%%%%%%%%%%%
\begin{center} \noindent \epsfxsize=0.9\columnwidth\epsffile{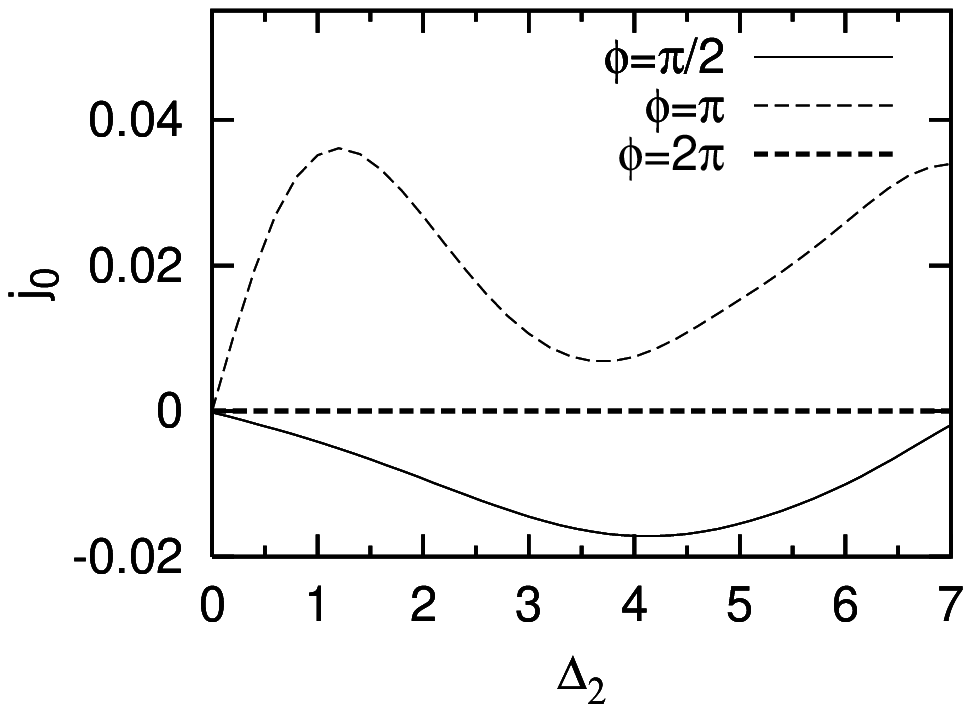}
\end{center}
\vskip-3mm\noindent{\footnotesize Fig. 6. The dc current flowing
through the double QD symmetrical system under the influence of
two external perturbations versus the amplitude $\Delta_2$ for the
phase difference $\phi=0=2\pi$ (thick broken line), $\phi=\pi$
(thin broken line) and $\phi=\pi/2$ (solid line). The other
parameters are $\omega_1=3$, $\Delta_1=3$, $\omega_2=6$,
$V_{12}=1$,
$\Gamma=0.2$, $\varepsilon_1 =\varepsilon_2=0$, $\mu_L=\mu_R=0$.}%
\vskip15pt
%%%%%%%%%%%%%%%%%%%%%%%%%%%%%%%%%%%%%%%%%%%%%%%%

A remaining question is how  the pump current depends on the phase
difference between both QD sites. In Fig.~7 we depict the dc
current flowing through a double QD symmetrical system
($\varepsilon_1 = \varepsilon_2=0$) as a function of the
(relative) phase $\phi$ for three values of the driving strength
of the second perturbation, i.e.\ $\Delta_2=3$ (solid line),
$\Delta_2=6$ (thick solid line) and $\Delta_2=9$ (broken line). In
case of zero phase difference, $\phi=0, 2\pi$,  the current is
zero although  two external perturbations are applied to our
system, cf. also Fig. 6. For  values of $\phi$ different from
$\phi=0, 2\pi$ a finite pump current flows through the double dot
system: Depending on the relative phase it can be either positive
or negative.
%%%%%%%%%%%%%%%%%%%%%%%%%%%%%%%%%%%%%%%%%%%%%%%
\begin{center} \noindent \epsfxsize=0.9\columnwidth\epsffile{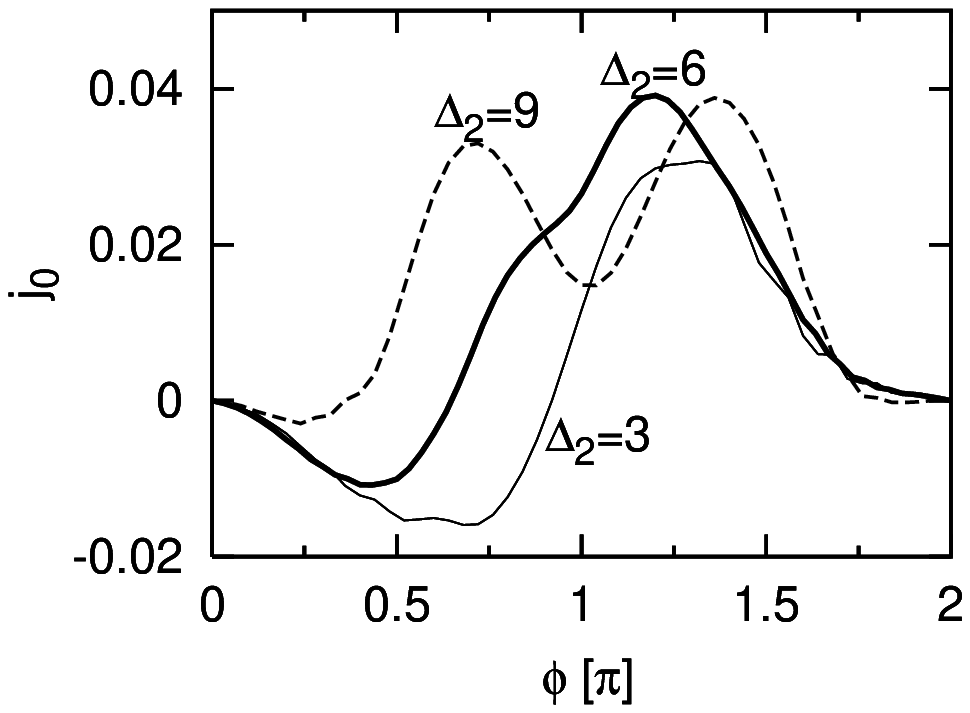}
\end{center}
\vskip-3mm\noindent{\footnotesize Fig. 7. The dc current  as a
function of the phase difference $\phi$ for two external
perturbations $\omega_1=3$, $\Delta_1=3$, $\omega_2=6$ and for
$\Delta_2=3$ (solid line), 6 (thick solid line) and 9 (broken
line). The other
parameters are: $V_{12}=1$, $\Gamma=0.2$, $\varepsilon_1 = \varepsilon_2=0$, $\mu_L=\mu_R=0$.}%
\vskip15pt
%%%%%%%%%%%%%%%%%%%%%%%%%%%%%%%%%%%%%%%%%%%%%%%%
\noindent  It is also possible to stop the pumping current all
together  for specific  values of $\phi$. Note, that for very
large amplitudes of external perturbation, $\Delta_2$, the current
takes on positive values for almost for all $\phi$, cf. the broken
line.

\section{\label{sec40}Conclusions}

Time-dependent electron transport through a quantum dot and double
quantum dot system in the presence of polychromatic perturbations has
been studied within the evolution operator method. The QD  has
been coupled with two electrodes and external, time-dependent energy perturbations have been
applied to the central region.

Analytical relations for the dc current flowing through the
system, Eq.~\ref{t03} and the charge accumulated on a QD,
Eq.~\ref{t04}, have been obtained for incommensurate external
perturbations. Also the analytical relations for the transmission
have been derived, Eq.~\ref{t01}  for commensurate frequencies and
for incommensurate frequencies, Eq.~\ref{t02}. It has been found
that sideband peaks appear for $\varepsilon_d=\sum_{i=1}^n \pm k_i
\omega_i$ where $k_i$ is an integer number and $n$ stands for the
total number of external perturbations. For the case of two
external perturbations this condition can be written as
$\varepsilon_d=\pm k_1 \omega_1 \pm k_2 \omega_2$ and e.g. for
$\omega_1=3$ and $\omega_2=8$ additional peaks appear for
$\varepsilon_d=2,5,...$, cf. Fig.~1. In the presence of  external
perturbations one can control
 the dc current by changing the amplitude strengths of the acting
perturbations, cf. Fig.~2. In the
presence of multiple external perturbations the current is
characterized by satellite peaks, cf.
Fig.~3 for two acting energy perturbations.

Moreover, the dc current obtained for commensurate frequencies, e.g. for bi-harmonic perturbations,
 strongly differs from that
one obtained for slightly different from commensurate frequencies,
cf. Fig.~4. Also the asymmetry in the transmission and the dc
current versus the quantum dot energy level  $\epsilon_d$  for
commensurate frequencies case is detected, see  in Fig. 5. This
asymmetry appears only for commensurate external perturbations
applied to the system and is related to a phase difference between
time-dependent perturbations. For a double QD system the
analytical formula for the transmission has been obtained for
incommensurate frequencies, Eq.~\ref{dqd03}. Also a electron
quantum pump based on a fully symmetrical double QD system has
been proposed in absence of   source-drain voltages and static
bias voltages
for which the pump current varies as a function of relative phase shift $\phi$, cf. Fig.~6 and Fig.~7.\\

%\section{\label{sec50}Acknowledgements}
\noindent\textbf{Acknowledgements}. This work has been partially
supported by  Grant No.\ N\,N202\, 1468\,33 of the Polish
Ministry of Science and Higher Education, the Alexander von
Humboldt Foundation (T.K.), the German-Israel-Foundation (GIF)
(P.H.) and the DFG priority program DFG-1243 ``quantum transport
at the molecular scale'' (P.H., S.K.). S.K. is supported by the
Ram\'on y Cajal program of the Spanish MICINN.

\end{multicols}
\end{document}